\documentclass[%
aps,%
preprint,%
showpacs,
nofootinbib,
superscriptaddress
]{revtex4}
\usepackage{epsfig}
\usepackage{amsmath}
\newcommand{\gtrsim}{ \mathop{}_{\textstyle \sim}^{\textstyle >} }
\newcommand{\lesssim}{ \mathop{}_{\textstyle \sim}^{\textstyle <} }
\newcommand{\wt}{\widetilde}

\def\be{\begin{equation}}
\def\ee{\end{equation}}
\def\ba{\begin{eqnarray}}
\def\ea{\end{eqnarray}}

\begin{document}

\preprint{CERN-PH-TH/2006-079}

\title{A possible symmetry of the  $\nu$MSM}

\author{Mikhail Shaposhnikov}
%\email{Mikhail.Shaposhnikov@epfl.ch}
\affiliation{
Institut de Th\'eorie des Ph\'enom\`enes Physiques,
Ecole Polytechnique F\'ed\'erale de Lausanne,
CH-1015 Lausanne, Switzerland}
\affiliation{Theory Division, CERN, CH-1211 Geneve 23, Switzerland}

%\date{\today}
\date{October 31, 2006}

%\vspace{4cm}

\begin{abstract}

To explain the dark matter and the baryon asymmetry of the Universe,
the parameters of the $\nu$MSM (an extension of the Minimal Standard
Model by three singlet neutrinos with  masses smaller than the
electroweak scale) must be fine-tuned: one of the masses should be in
the ${\cal O} (10)$ keV region to provide a candidate for the
dark-matter particle, while two other masses must be almost the same
to enhance the CP-violating effects in the sterile neutrino
oscillations leading to the baryon asymmetry. We show that a
specifically defined global  lepton-number symmetry, broken on the
level of ${\cal O} (10^{-4})$ leads to the required pattern of
sterile neutrino masses being consistent with the data on neutrino
oscillations. Moreover, the existence of this symmetry allows to fix
the flavour structure of couplings of singlet fermions to the
particles of the Standard Model and indicates that their masses are
likely to be smaller than ${\cal O} (1)$ GeV, opening a possibility of
their search in decays of charmed, beauty and even $K$ or
$\pi$-mesons.  

\end{abstract}


\maketitle

%%%%%%%%%%%%%%%%%%%%%%%%%%%%%%%%%%%%%%%%%%%%%%%%%%%%%%%%%%%%%%%%%%%%%%%%
% Introduction
%%%%%%%%%%%%%%%%%%%%%%%%%%%%%%%%%%%%%%%%%%%%%%%%%%%%%%%%%%%%%%%%%%%%%%%%
\section{Introduction}
There is compelling evidence that the Minimal Standard Model (MSM) of
strong and electroweak interactions is not complete. There are
several {\em experimental} and {\em observational} facts that cannot
be explained by the MSM. These are neutrino oscillations, the
presence of dark matter in the Universe, the baryon asymmetry of the
Universe, its flatness, and the existence of cosmological
perturbations necessary for structure formation. Indeed, in the MSM
neutrinos are strictly massless and do not oscillate. The MSM does
not have any candidate for non-baryonic dark matter. Moreover, with
the present experimental limit on the Higgs mass, the
high-temperature phase transition, required for electroweak
baryogenesis, is absent. Finally, the couplings of the single scalar
field of the MSM are too large for the Higgs boson to play the role
of the inflaton.

In this paper we will assume that a simple renormalizable extension
of the MSM by three singlet right-handed neutrinos and by a real
scalar field (inflaton) with masses smaller than the electroweak
scale is a correct effective theory up to some high-energy scale,
which may be as large as the Planck scale
\cite{Asaka:2005an,Asaka:2005pn,Shaposhnikov:2006xi}.  Following
\cite{Asaka:2005an,Asaka:2005pn}, we will call this model ``the
$\nu$MSM", underlying the fact that it is the extension of the MSM in
the neutrino sector. We will use the same name for a theory with
inflaton \cite{Shaposhnikov:2006xi}. Contrary to Grand Unified
Theories, the $\nu$MSM does not have any internal hierarchy problem,
simply because it is a theory with a single mass scale. Moreover, as
the energy behaviour of the gauge couplings in this theory is the
same as in the MSM, the absence of gauge-coupling unification in it
indicates that there may be no grand unification, in accordance with
our assumption of the validity of this theory up to the Planck scale.
As well as the MSM, the $\nu$MSM does not provide any explanation why
the weak scale is much smaller than the Planck scale.

As we demonstrated in \cite{Asaka:2005an,Asaka:2005pn}, the $\nu$MSM 
can explain simultaneously dark matter and baryon asymmetry of the
Universe being consistent with neutrino masses and mixings observed
experimentally\footnote {We do not include here the LSND
anomaly~\cite{Aguilar:2001ty}, which will be tested in MiniBooNE
experiment ~\cite{MiniBooNE}.}. Moreover, in
\cite{Shaposhnikov:2006xi} we have shown that inclusion of an
inflaton with scale-invariant couplings to the fields of the $\nu$MSM
allows us to have inflation and provides a common source for
electroweak symmetry breaking and Majorana neutrino masses of singlet
fermions -- sterile neutrinos. The role of the dark matter is played
by the lightest sterile neutrino with mass $m_s$ in the  keV range,
which can be created in active--sterile neutrino oscillations
\cite{Dodelson:1993je} or through the coupling to the inflaton
\cite{Shaposhnikov:2006xi}. In the latter case, the dark matter
sterile neutrino can satisfy  \cite{Asaka:2006ek,Shaposhnikov:2006xi}
all astrophysical and cosmological constraints on its mass and mixing
angle with an active neutrino  \cite{Boyarsky:2005us,Boyarsky:2006zi,
Boyarsky:2006fg,Riemer-Sorensen:2006fh,Seljak:2006qw,Viel:2006kd,Watson:2006qb}.
In addition,  the coherent oscillations of two other, almost
degenerate, sterile neutrinos  lead to the creation of baryon
asymmetry of the Universe \cite{Asaka:2005pn}  through the splitting
of the lepton number between active and sterile neutrinos
\cite{Akhmedov:1998qx} and electroweak sphalerons
\cite{Kuzmin:1985mm}. For other astrophysical applications of sterile
neutrinos see \cite{Kusenko:1997sp,Biermann:2006bu,Stasielak:2006br}.

To explain simultaneously the dark matter and baryon asymmetry of the
Universe, the parameters of the $\nu$MSM must be highly fine-tuned
\cite{Asaka:2005an,Asaka:2005pn}.  If $m_s$, the mass of the lightest
sterile neutrino $N_1$, playing the role of dark matter, is in the
keV range (this mass interval is suggested, for example, by
observations of dwarf galaxies \cite{Goerdt:2006rw} and by arguments
coming from structure formation \cite{Bode:2000gq}), it must be about
five orders of magnitude smaller than the masses of the heavier
sterile neutrinos, $N_2$ and $N_3$.  The Yukawa coupling of $N_1$ to
left-handed leptons ($f_1 \lesssim 10^{-12}$) must be at least five
orders of magnitude smaller than the similar coupling of  $N_2$ and
$N_3$ (these constraint are coming from the Big Bang Nucleosynthesis
(BBN)). Moreover, the heavier sterile neutrinos $N_2$ and $N_3$ must
be degenerate in mass at least in one part in $10^5$. This looks
rather bizarre. However, these fine-tunings can be taken as an
indication of some new (slightly broken) symmetry, which the $\nu$MSM
must satisfy and which follows from some unknown underlying physics,
related, perhaps, to the Planck scale. This paper is an attempt to
reveal such a symmetry. We will show that the above-mentioned fine
tunings can be explained by the approximate conservation of a
specifically defined lepton number, explicitly broken by Majorana
mass terms and Yukawa coupling constants.

We would like to stress that this work does not try to shed any light
on the structure of the {\it active} neutrino mass matrix. For
example, we have nothing to say about the smallness of the mixing
angle $\theta_{13}$ or why the mixing angle $\theta_{23}$ is so close
to maximal. Our aim is different: we want to understand, on the basis
of the available experimental data on neutrino oscillations and on
input from cosmology, described above, what are the properties of
heavier singlet fermions, and, in particular, what are the values of
their masses and couplings to left-handed leptons.  We will show that
if the breaking of the lepton number symmetry (that makes the heavier
sterile neutrinos be degenerate and the dark matter neutrino be
light) is tiny, then the masses of singlet fermions should be 
smaller than  $1$ GeV. Thus, these particles can be searched for in
decays of charmed and beauty mesons, $\tau$-lepton and, possibly $K$
or $\pi$-mesons (in the latter case a number of experimental and BBN
constraints already exist). In addition, we show that the ratios of
the  Yukawa couplings of sterile neutrinos to the Higgs and SM
leptons can be found from the data on active neutrino mass matrix.  

The paper is organized as follows. In Section 2 we will discuss the
requisite symmetry of the $\nu$MSM that can explain the 
above-mentioned fine tunings and construct a symmetric $\nu$MSM. In
the third Section we introduce the global symmetry breaking and
analyse the active neutrino mixing matrix; in the fourth we will
discuss the baryon asymmetry of the Universe and dark matter in the
nearly symmetric $\nu$MSM. In Section 5 we discuss the  values of the
masses of heavy neutral leptons and consider the Yukawa couplings of
singlet fermions to the Higgs and left-handed leptons. We conclude in
Section 6.

%%%%%%%%%%%%%%%%%%%%%%%%%%%%%%%%%%%%%%%%%%%%%%%%%%%%%%%%%%%%%%%%%%%%%%%%
% Chapter 2
%%%%%%%%%%%%%%%%%%%%%%%%%%%%%%%%%%%%%%%%%%%%%%%%%%%%%%%%%%%%%%%%%%%%%%%%
\section{The symmetric $\nu$MSM} 
The Lagrangian of the $\nu$MSM is 
\begin{eqnarray}
  {\cal L}_{\nu\rm{MSM}}
  = {\cal L}_{\rm{MSM}} + \bar N_I i \partial_\mu \gamma^\mu N_I
  - F_{\alpha I} \,  \bar L_\alpha N_I \Phi
  - \frac{M_{IJ}}{2} \; \bar {N_I{}^c} N_J + \rm{h.c.} \,,
  \label{lagr}
\end{eqnarray}
where $\Phi$ and $L_\alpha$ ($\alpha=e,\mu,\tau$) are respectively
the Higgs and lepton doublets, $F$ is a matrix of Yukawa coupling
constants, and $M_I$ are the Majorana masses. By the field
redefinitions, the Majorana mass matrix can be taken to be real and
diagonal\footnote{In \cite{Shaposhnikov:2006xi} the Majorana masses
were coming from interaction with the inflaton field. In this paper
we replace the inflaton field by its vacuum expectation value, since
the discussion of symmetry properties does not require any analysis
of inflaton dynamics.}, $M_{IJ}=M_I\delta_{IJ}$. Without loss of
generality we choose the matrix of Yukawa couplings of left-handed
doublets  $L_\alpha$ and right-handed charged leptons $E_\alpha$ to
be real and diagonal.

Suppose now that all fine tunings of the $\nu$MSM, necessary for the
production of dark matter and baryon asymmetry of the Universe, are
{\em exactly} satisfied. (To distinguish the fields in this case from
those appearing in (\ref{lagr}), we will use a ``tilde" in the
notation.) Namely, let us require that $\wt N_2$ and $\wt N_3$ be
{\em exactly} degenerate in mass, that $\wt N_1$ be {\em exactly}
massless and {\em not to interact at all} with left-handed doublets.
What kind of global symmetry can lead to this pattern automatically? 

As for degeneracy between $\wt N_2$ and $\wt N_3$, the unique choice
(others are equivalent to this one after linear transformations of
fields) is the U(1) symmetry:
\be
\wt N_2 \rightarrow e^{-i\alpha_L} \wt N_2,
~~~ \wt N_3 \rightarrow e^{i\alpha_L} \wt N_3~,
\ee
whereas for $\wt N_1$ this is simply a chiral symmetry 
\be
\wt N_1 \rightarrow e^{i\beta_L} \wt N_1.
\ee
One can say that  $\wt N_2$ and $\wt N_3$ form a Dirac spinor $\Psi =
\wt N_2 + \wt N_3^c$, or, in other words, $\wt N_2$ and $\wt N_3$
represent particle and antiparticle. As a result, the free Lagrangian
for sterile neutrinos is
\be
L_{\rm{free}}= \bar {\wt N_I} i \partial_\mu \gamma^\mu \wt N_I 
- M \bar {\wt N_2^c} \wt N_3~.
\label{free}
\ee
It is invariant under $U(1)_{23}\times U(1)_1$ symmetry.  One can
equally say that there is just one $U(1)_L$ symmetry and that the
quantum numbers of $\wt N_I$ with respect to this symmetry are
($q,-1,1$) for $\wt N_1,~\wt N_2$ and $\wt N_3$ respectively, and
that $q \neq 0$, since for $q=0$ a Majorana mass term for $\wt N_1$
is allowed. Though for $q=\pm 1$ the mass mixing terms between $\wt
N_{2,3}$ and  $\wt N_1$ are admitted, the theory still has one
massless state in the sterile neutrino sector. We will call the
quantum number associated with this $U(1)_L$ symmetry ``the lepton
number" for reasons that will become obvious below.  The parameter
$M$ is nothing but the mass of $\wt N_{2,3}$. In addition to $U(1)_L$
symmetry specified above the Lagrangian ${\cal L}_{\nu\rm{MSM}}
+ L_{\rm{free}}$ is invariant under $U(1)_e\times U(1)_\mu \times
U(1)_\tau$ symmetry corresponding to separate conservation of
$e,~\mu$ and $\tau$ leptonic numbers. At this stage the total global
symmetry in the leptonic sector is $U(1)^4$.

Now, we would like to construct an interaction, consistent with our
$U(1)_L$ symmetry, between $\wt N_{2,3}$ and $L_\alpha$, but forbid the
interaction between $\wt N_1$ and $L_\alpha$. The assumption that
{\em all} leptonic doublets are allowed to interact with sterile
neutrinos $\wt N_{2,3}$ and with right-handed charged leptons leads
us to four inequivalent charge assignments for active leptonic
flavours: $(1,1,1),~(1,1,-1),~(1,-1,1)$ and $(-1,1,1)$. Then, the
interactions between $\wt N_1$ and $L_\alpha$ are not allowed for $q
\neq \pm 1$. So, we have to consider four different Yukawa matrices
$F_0$: 
\begin{eqnarray}
\label{11}\rm{Model~~ I:}~~~
F_0=\left(
   \begin{array}{c c c}
     0& h_{12} & 0  \\
     0& h_{22} & 0 \\
     0& h_{32} & 0 
   \end{array}
  \right)~,~~~
\label{22}\rm{Model~~ II:}~~~
F_0=\left(
   \begin{array}{c c c}
     0&0 & h_{13} \\
     0&0 & h_{23} \\
     0&h_{32} & 0 
   \end{array}
  \right)~,
    \end{eqnarray}
\begin{eqnarray}  
\label{33}\rm{Model~~ III:}~~~
F_0=\left(
   \begin{array}{c c c}
     0&0 & h_{13} \\
     0&h_{22}& 0  \\
     0&0 & h_{33} 
   \end{array}
  \right)~,~~~
\label{44}\rm{Model~~ IV:}~~~
F_0=\left(
   \begin{array}{c c c}
     0&h_{12}& 0  \\
     0&0 & h_{23} \\
     0&0 & h_{33} 
   \end{array}
  \right)~.
\end{eqnarray}
where $h_{ij}$ can be taken to be real. In any of these models the 
total global symmetry in the leptonic sector is broken explicitly
down to $U(1)_L$.

To summarize, we achieved the initial goal: the theory is invariant
under $U(1)_L$ global symmetry, which guarantees the degeneracy
between $\wt N_2$ and $\wt N_3$, the absence of mass for $\wt N_1$, and
the absence of interactions between $\wt N_1$ and $L_\alpha$.

To choose the charge assignment that could best fit the data on
neutrino oscillations, let us consider the active neutrino mass
matrix, which we define following ref. \cite{Strumia:2005tc}:
\be
M_\nu= V^*\cdot {\rm diag}(m_1,m_2e^{2i\delta_1},m_3e^{2i\delta_2})
\cdot V^\dagger~,
\label{act}
\ee
with $V = R(\theta_{23}) \rm{diag}(1,e^{i\delta_3},1) R(\theta_{13})
R(\theta_{12})$ the active neutrino mixing matrix \cite{numix}, and
choose for normal hierarchy $m_1<m_2<m_3$ and for inverted hierarchy
$m_3<m_1<m_2$. All active neutrino masses are taken to be positive.

One can easily see that for models II--IV one active neutrino is
massless while the other two are degenerate. So, only the inverted
hierarchy of active neutrino masses can be realized. As for the
mixing angles in the active neutrino mass matrix,  one finds for 
Model II: $\theta_{12}=\pi/4,~\theta_{23}=\pi/2,~\theta_{13}=-\arctan
(h_{23}/h_{13})$; for Model III:
$\theta_{12}=\pi/4,~\theta_{23}=0,~\theta_{13}=-\arctan
(h_{33}/h_{13})$, and for Model IV :
$\theta_{12}=\pi/4,~\theta_{13}=0,~\theta_{23}=-\arctan
(h_{33}/h_{23})$, with a common choice of $\delta_3=0,
\delta_1=\pi/2$. Clearly,  Models II and III cannot fit the data on
neutrino oscillations, whereas  Model IV is in fact quite close to
the data, provided $h_{23} \simeq h_{33}$. This charge assignment
corresponds to the so called $(L_e-L_\mu-L_\tau)$ symmetry
\cite{Petcov:1982ya,Barbieri:1998mq}. In the simplest forms of the
breaking of this symmetry, the deviation of the solar mixing angle
from its maximal value is too small, and this scheme is disfavoured
by the data \cite{He:2002rv} (see, however, \cite{Altarelli:2005pj}).

We will choose, therefore, Model I as our starting point. The
$U(1)_L$ global symmetry of this model is uniquely determined by the
cosmological requirements formulated in the Introduction and by the
data on neutrino oscillations. In this theory all three active
neutrino masses are necessarily equal to zero and the mixing angles
cannot be determined, so that both masses and angles are fixed by the
breaking of the $U(1)_L$ symmetry.

%%%%%%%%%%%%%%%%%%%%%%%%%%%%%%%%%%%%%%%%%%%%%%%%%%%%%%%%%%%%%%%%%%%%%%%%
% Chapter 3
%%%%%%%%%%%%%%%%%%%%%%%%%%%%%%%%%%%%%%%%%%%%%%%%%%%%%%%%%%%%%%%%%%%%%%%%
\section{Symmetry breaking and active neutrino masses and mixings}
The $U(1)_L$ symmetric theory constructed above contradicts 
experiment. Indeed, all active neutrino masses are zero. There is no
candidate for dark matter particle simply because $\wt N_2$ and $\wt
N_3$ are not stable and $\wt N_1$ is massless. The model does not
contain any CP-violating phase and the baryon asymmetry cannot be
produced. Therefore, the $U(1)_L$ symmetry cannot be exact. We will
assume that this symmetry is slightly broken both by the Majorana
mass terms and by the Yukawa interactions. So, we add to the
symmetric Lagrangian  the following symmetry-breaking terms
\be
L_{\rm{breaking}}=\Delta L_{\rm{mass}} + \Delta L_{\rm{Yukawa}}~,
\ee
where 
\be
\Delta L_{\rm{mass}}= -\frac{1}{2} \; \bar {\wt N^c}\Delta M \wt N~,
\ee
and 
\be
\Delta L_{\rm{Yukawa}}= - \bar L \Delta F \wt N \Phi~.
\ee
Explicitly,
\be
\Delta M= \left(
   \begin{array}{c c c}
      m_{11}e^{i\alpha}& m_{12}                  & m_{13}\\
     m_{12}                &  m_{22}e^{i\beta}   & 0\\
     m_{13}           & 0             & m_{33}e^{i\gamma}
   \end{array}
  \right)~.
\ee
Here all $m_{ij} \ll M$ can be taken to be real, $\alpha,~\beta$,
and $\gamma$ are three Majorana CP-violating phases.

The Yukawa part is taken to be 
\be
\Delta F = \left(
   \begin{array}{c c c}
    h_{11}& 0  &  h_{13} \\
    h_{21}& 0  &  h_{23} \\
    h_{31}& 0  &  h_{33}
   \end{array}
  \right)~,
  \label{deltaF}
\ee
where $h_{i1}\ll h_{k2},~ h_{i3}\ll h_{k2}$ are in general complex,
containing 3 physical CP-breaking phases.

We do not have at hand any deep theory guideline  to choose this or
that pattern of symmetry breaking. Therefore, for the choice of
parameters, we will be using phenomenological considerations, related
to dark matter and the baryon asymmetry of the Universe
\cite{Asaka:2005an,Asaka:2005pn}.  To characterize the measure of the
$U(1)_L$ symmetry breaking\footnote{Of course, the symmetry breaking
Yukawa coupling constants and masses are subject to renormalization
and do depend on the normalization scale $\mu$. This dependence,
however, is completely negligible for $\mu < M_{Pl} \sim 10^{19}$ GeV
due to the smallness of Yukawa couplings in the $\nu$MSM and can be
safely omitted.}, we introduce a small parameter $\epsilon =F_3/F \ll
1$, where $F_i^2 = [h^\dagger h]_{ii}$, $h=F_0+\Delta F$, and
$F_2\equiv F$. 

Let us consider first the masses of sterile neutrinos in a theory
with symmetry breaking. As for the lightest sterile neutrino, playing
the role of dark matter particle, we get, expanding in powers of
$m_{ij}$:
\be
m_s=M_1 \simeq \left(m_{11} - \frac{2m_{12}m_{13} \cos\alpha}{M}\right)~.
\ee
For the Majorana mass square difference of the heavier neutrinos,
essential for baryogenesis\footnote{We disregard here the
contribution to the physical masses of heavier Majorana neutrinos,
coming from the electroweak symmetry breaking, which is of the order
of active neutrino masses, $\sim 0.05$ eV.}, we find
\ba
M_2^2-M_3^2\simeq 2\left[4 m_{12} m_{13}\,\left(m_{12}m_{13}
+
M m_{33}\cos \gamma  + 
     M m_{22}\cos\beta \right) 
     + \right.\\ 
\left.
   + M^2 \left({m_{33}}^2 + {m_{22}}^2 + 
     2m_{33}m_{22} \cos(\gamma  + \beta) \right)\right]^{1/2}~.
\ea
If, for example, all $m_{ij} \propto \epsilon^n M$, we obtain
parametrically $|M_2-M_3| \sim M_1$. For a somewhat different
breaking pattern: $m_{11}\propto m_{12}\propto m_{13} \propto
\epsilon^n M,~~m_{33}\propto m_{22}\propto \epsilon^{2n} M$ we
obtain $M_2^2-M_3^2 \sim M_1^2$. 

The mass eigenstates ($N_I$ without tilde) are related to $\wt N_I$ by
the unitary transformation,
\be
\wt N = U_R N~.
\ee
Different elements of the matrix $U_R$ are, parametrically:
\ba
\nonumber
\left[ U_R\right]_{11} \simeq 1,~~
[U_R]_{1i} \sim \frac{m_{1i}}{M},~~
[U_R]_{j1} \sim \frac{m_{1j}}{M}~,\\
\nonumber
\\
\left[ U_R\right]_{ij} \simeq \frac{e^{i\phi_0}}{\sqrt{2}}
\left(
   \begin{array}{c c }
      e^{i\phi_1} &  e^{i\phi_2}\\
      - e^{-i\phi_2}&  e^{-i\phi_1}
   \end{array}
  \right)~,
  \label{UR}
\ea
where $i,j=2,3$, and the phases $\phi_k$ are related to
$\alpha,~\beta,~\gamma$ and $m_{kl}$ through some expressions, which
we do not present here because of their complexity and of the fact
that they will not be used in what follows. In is clear from eq.
(\ref{UR}) that for small $\epsilon$ the heavy sterile neutrino mass
eigenstates have almost identical Yukawa couplings to active
neutrinos.

Now, let us turn to active neutrino masses. Their values can be found
from the general see-saw formula:
\be
M_\nu = - \sum_I F^* \frac{v^2}{M_I} F^\dagger~.
\label{seesaw}
\ee
The computation is simplified considerably by noting 
\cite{Asaka:2005an,Asaka:2005pn,Boyarsky:2006jm} that the coupling of
the lightest sterile neutrino to active fermions must be very small
to avoid the constrains coming from the over-closing of the Universe
and from X-rays observations. This coupling contains three different
contributions, the direct one, $\sim h_{i1}$, and two induced, coming
from the mixing defined in (\ref{UR}): $\sim h_{i3}\frac{m_{13}}{M}$
and $\sim h_{i2}\frac{m_{12}}{M}$. Unless a miraculous cancellation
of different contributions takes place, all these terms must be smaller
than ${\cal O}(10^{-12})$. Thus, in the leading order we find that
the smallest active neutrino mass is given by:
\be
m_1= \frac{F_1^2 v^2}{m_{11}}~.
\ee
To find two other masses we expand eq. (\ref{seesaw}) with respect to
small parameters $h_{i1}/h_{k2}\ll 1,~~ h_{i3}/h_{k2}\ll
1,~~m_{ij}\ll M$.  In the leading approximation, the active neutrino
mass matrix has a simple form
\be
[M_\nu]_{\alpha\beta} = \frac{v^2}{M}\left({\tilde h}_{\beta 3}{
h}_{\alpha 2}+{\tilde h}_{\alpha 3}{h}_{\beta 2}\right)~,
  \label{amass}
\ee
where
\be
{\tilde h}_{\beta 3}=h_{\beta 3} -\frac{1}{2} \frac{m_{33}}{M}
h_{\beta 2}~~.
\nonumber
\ee
The matrix (\ref{amass}) has one zero eigenvalue, corresponding to
the lightest active neutrino, so one has to choose $m_1=0,~ m_2 = 
m_{\rm{sol}}$, and $m_3 =  m_{\rm{atm}}$, for the case of normal
hierarchy, or $m_3=0,~ m_1 =  m_{\rm{atm}}-\Delta m/2$, and  $m_2 = 
m_{\rm{atm}}+\Delta m/2$, for the case of inverted hierarchy (here
$\Delta m=m^2_{\rm{sol}}/m_{\rm{atm}}$). The two non-zero eigenvalues of
$M_\nu M_\nu^\dagger$ give for other active neutrino masses:
\be
m_{2,3}= \frac{v^2}{M}\left({F}_2 {\tilde F}_3 \pm
|h^\dagger {\tilde h}|_{23}\right)~,
\ee
where $v=174$ GeV is the vacuum expectation value of the Higgs field,
${\tilde F_3}$ is constructed from ${\tilde h}_{\beta 3}$ in full analogy with
$F_3$.
Replacing $F_2$ by $F$ and ${\tilde F}_3$ by $\epsilon
{\tilde F}$ we then have:  
\ba
\nonumber
2 {F}^2 v^2 \epsilon/M \simeq m_{\rm{atm}}~
[\rm{normal~hierarchy}]~,\\
{F}^2 v^2 \epsilon/M \simeq m_{\rm{atm}}~
[\rm{inverted~hierarchy}]~,
\label{Ffix}
\ea
where $m_{\rm{atm}}=\sqrt{\Delta m^2_{\rm{atm}}}\simeq 0.05$ eV and
$m_{\rm{sol}}=\sqrt{\Delta m^2_{\rm{sol}}}\simeq 0.01$ eV.  If the
``vectors" ${h}_{i2}$ and ${\tilde h}_{i3}$ are almost
parallel, the normal hierarchy of active neutrino masses is realized
(this happens automatically if $\frac{m_{33}}{M} h_{\beta 2}\gg
h_{\beta 3}$); if they are almost orthogonal, the inverted hierarchy
is achieved. 

Let us now discuss whether this matrix can describe the observed
neutrino mixings parametrized by $V$. Suppose that $M$  is fixed.
Then the question is: Can eq. (\ref{amass}), considered as a set of
six equations for six complex numbers ${h}_{i2},~{\tilde
h}_{i3}$ always be solved? The answer to this question is always
affirmative; moreover, the solution is not unique and can be
parametrized by an arbitrary complex number $z$. This can be seen as
follows: the matrix $M_\nu$ does not change if we replace ${
h}_{i2}$ by $z { h}_{i2}$ and ${\tilde h}_{i3}$ by  ${\tilde
h}_{i3}/z$.  Now, since we defined $\epsilon$ from ${\tilde
F}_3=\epsilon F_2$ we have $\epsilon =|z|$. In other words,
from the point of view of the data on neutrino oscillations, the
measure of the breaking of the $U(1)_L$ symmetry cannot be determined
and can be arbitrarily small. As for the well-known peculiarities of
the active neutrino mixing matrix (smallness of the $\theta_{13}$
mixing angle and the maximal value of $\theta_{23}$), they can always
be achieved with a certain choice of $h_{ij}$.

In the next section we will discuss the constraints on $\epsilon$
coming from the baryon asymmetry of the Universe and from dark
matter.

%%%%%%%%%%%%%%%%%%%%%%%%%%%%%%%%%%%%%%%%%%%%%%%%%%%%%%%%%%%%%%%%%%%%%%%%
% Chapter 4
%%%%%%%%%%%%%%%%%%%%%%%%%%%%%%%%%%%%%%%%%%%%%%%%%%%%%%%%%%%%%%%%%%%%%%%%
\section{Baryon asymmetry and dark matter}
We will require that the theory with global $U(1)_L$  symmetry, which
is slightly broken,  constructed in the previous section, gives the
baryon asymmetry of the Universe. This will give us a lower bound on
the parameter $\epsilon$, for the following reasons:  first, if we
decrease $\epsilon$, the Yukawa coupling $F$, required for
explanation of the atmospheric neutrino mass scale (\ref{Ffix}), must
increase (for $M$ fixed). If $F$ is too large, the sterile neutrinos
$N_{2,3}$ will equilibrate well before the electroweak scale. In this
case the production of baryon asymmetry is suppressed
\cite{Asaka:2005pn}.  Second, in the limit $\epsilon \to 0$, the
CP-violating effects in the neutrino sector go away. This fact,
though, is less important than the first one.

To estimate these effects, consider the equilibration rate of
$N_{1,2}$, which in our notation is given by 
\cite{Akhmedov:1998qx,Asaka:2005pn} 
\be 
\Gamma_N \simeq \frac{F^2 T}{8} \sin\phi~, 
\ee 
where $\sin\phi \simeq 0.02$. A deviation of the concentration of
$N_{1,2}$ from the equilibrium one at temperature $T$ is then
proportional to 
\be
\delta n \sim \exp\left(-\frac{\Gamma_N M_0}{T^2}\right)~,
\label{noneq}
\ee
where $M_0 \simeq 7 \times 10^{17}$ GeV  appears in the
time--temperature relation, $t = \frac{M_0}{2 T^2}$.

If  
\be
F\lesssim 3\times 10^{-7},
\label{Fweak}
\ee
then $N_{2,3}$ do not equilibrate till they reach the temperature
$T_W \simeq 130$--$190$ GeV, corresponding to the freeze out of the
sphaleron transitions \cite{Burnier:2005hp}, and the baryon asymmetry
of the Universe is given by the general eqs. (22) and (29) of 
\cite{Asaka:2005pn}\footnote{Note that the matrix of the Yukawa
couplings $F$ introduced in \cite{Asaka:2005pn} was given in the
Majorana mass eigenstate basis. It is equal to $(F_0+\Delta F )U_R$
in the notation of the present paper.}: 
\be
\frac{n_B}{s} \sim 7 \times 10^{-4} \rm{Tr}\Delta_N\vert_{T_W}~, 
\ee
where the asymmetry in the sterile neutrino sector
$\Delta_N\vert_{T_W}$ is given by 
\cite{Asaka:2005pn} :
\be
\Delta_N\vert_{T_W} \sim a
  \frac{\pi^{\frac 32} \, \sin^3 \phi }
  { 96 \cdot 3^{\frac{1}{3}} \Gamma (\frac{5}{6})} \, 
  \frac{\epsilon F^6 M_0{}^{\frac{7}{3}}}
  {T_W(\Delta M_{32}^2)^{\frac{2}{3}}} \,.
  \label{cp}
\ee
Here $a$ is a function of CP-violating phases that can be of the
order of $1$. The asymmetry $\Delta_N\vert_{T_W}$ is maximal and can
be of the order of $1$ provided the typical temperature of the lepton
asymmetry generation, determined from the condition
\be
\frac{|M_2^2-M_3^2|M_0}{T^3} \sim 1~,
\ee
coincides with the temperature at which sphaleron processes are
switched off; this happens if  $|M_2^2-M_3^2|\sim T_W^3/M_0 \simeq
(2~\rm{keV})^2$. Quite interestingly, a keV scale in the mass square
differences of sterile neutrinos extremizes the baryon asymmetry. 

If the lepton asymmetry production has a resonance character, a
constraint on $F$ from above can be somewhat relaxed. Indeed, if
$N_{2,3}$ are close to thermal equilibrium, the result (\ref{cp})
will be suppressed by $\delta n$ defined in (\ref{noneq}), and
$\delta n$ as small as $10^{-7}$ could still lead to the observed
baryon asymmetry. This gives $F \lesssim 1.2\times 10^{-6}$, which is
about $4$ times weaker than the requirement that reactions with
$N_{2,3}$  be out of thermal equilibrium. This constraint, together
with  (\ref{Ffix}), leads to 
\be
\epsilon \gtrsim 6\times 10^{-4} (M/\rm{GeV})
\label{cN}
\ee
for the normal hierarchy  and to  
\be
\epsilon \gtrsim 1.2 \times 10^{-3} (M/\rm{GeV})
\label{cI}
\ee
for the inverted hierarchy. If the value of $F$ from eq. (\ref{Fweak})
is taken, the values of $\epsilon$ have to be larger than those in
(\ref{cN},\ref{cI}) by a factor of $16$. 

Let us discuss now the constraints on the symmetry breaking pattern
coming from dark matter. Since the coupling of dark matter sterile
neutrinos to active neutrinos must be sufficiently suppressed, one
should have $m_{12}\propto \epsilon^p M$, $m_{13}\propto \epsilon^q
M$ and  $h_{i1} \propto \epsilon^l h_{i2}$, with $p \geq 2,~~q \geq
1$ and $l \geq 2$. A dark matter neutrino in the keV region would
correspond to $m_{11} \propto \epsilon^2 M$, for $M \sim 1$ GeV. With
this pattern of symmetry breaking an ${\cal O}(10)$ keV dark matter
neutrino could potentially be observed in astrophysical X-ray
observations. However,  with $\epsilon \sim 10^{-3}$ and  $p \geq
3,~~q \geq 2$ and $l \geq 3$,  the decay rate $N_1\to\nu\gamma$ would
be highly suppressed and observing a line in the X-ray background
corresponding to radiative decays of an ${\cal O}(10)$ keV sterile
neutrino seems to be very difficult, if not impossible. As for
sterile neutrino production, it could be in this case entirely 
related to interactions with the inflaton \cite{Shaposhnikov:2006xi}.

%%%%%%%%%%%%%%%%%%%%%%%%%%%%%%%%%%%%%%%%%%%%%%%%%%%%%%%%%%%%%%%%%%%%%%%%
% Chapter 5
%%%%%%%%%%%%%%%%%%%%%%%%%%%%%%%%%%%%%%%%%%%%%%%%%%%%%%%%%%%%%%%%%%%%%%%%
\section{Masses and couplings of singlet fermions}

Let us discuss first possible values of masses of singlet fermions.
As follows from  Eqns. (\ref{cN},\ref{cI}), the smaller the mass $M$
of the singlet fermion, the more exact the $U(1)_L$ symmetry can be.
So, the requirement of ``naturalness" of the required sterile
neutrino mass pattern favours light singlet fermions.  However, very
small masses are not allowed by the number of reasons. Indeed, the
reducing of $M$ requires the simultaneous decrease of the Yukawa
coupling $F$ (see eq. (\ref{Ffix})), which results in the decrease of
the baryon asymmetry of the Universe, see eq. (\ref{cp}). This
constraint, however, is rather weak and allows to have $M$ as small 
as few MeV. However,  the lifetime of very light sterile neutrinos
may be too large $\tau \gtrsim 10^{-2}$ s and thus spoil the
predictions of BBN \cite{Dolgov:2000jw,Dolgov:2000pj}.

To discuss the BBN constraints, let us consider first the $\nu$MSM
{\it without inflaton}. In this theory singlet neutrinos decay trough
the mixing with charged leptons and active neutrinos and the relevant
limits can be extracted from  \cite{Dolgov:2000jw,Dolgov:2000pj}.
These limits  can be contrasted with experimental results of
\cite{Bernardi:1987ek}, what gives a lower bound on the mass of a
singlet fermion $M \gtrsim 140$ MeV \cite{Kusenko:2004qc,Gorbunov}.

Now, if light inflaton $\chi$ with mass $m_\chi$ is added to the
theory, as in \cite{Shaposhnikov:2006xi}, singlet fermions acquire a
new fast decay mode $N_{2,3} \rightarrow \nu \chi$, provided $M >
m_\chi$, and the BBN constraints disappear if $M>$ few MeV. In other
words,  singlet fermions, responsible for baryon asymmetry of the
Universe, can be searched for at K2K, MiniBooNe and MINOS experiments
in $K$ and even in $\pi$-meson decays.  A discussion of different
signatures of relatively light sterile neutrinos in a theory without
inflaton can be found in \cite{Kusenko:2004qc}.

Neutral fermions with masses $M > 400$ MeV can be created in decays
of $D$ and $B$-mesons or $\tau$-lepton. The possibilities for their
experimental search will be discussed elsewhere \cite{Gorbunov}
(existing experimental and BBN constraints for these masses are too
weak, see
\cite{Bernardi:1987ek,Dolgov:2000jw,Dolgov:2000pj,Astier:2001ck,Kusenko:2004qc}). 

With a choice of $M$ of, say, $100$ MeV the measure of the breaking of
the leptonic number symmetry can be as small as $\epsilon \sim
10^{-4}$. For larger masses the breaking of $U(1)_L$  must be
stronger, see (\ref{cN},\ref{cI}).

Now, let us turn to interactions of singlet fermions. The ratios of
the Yukawa couplings $h_{i2}$ can be expressed through the elements
of the active neutrino mixing matrix. A particularly simple expression
can be derived for the case $\theta_{13}=0,~\theta_{23}=\pi/4$, which
is in agreement with the experimental data. For normal hierarchy four
solutions are possible:
\be
|h_{12}|:|h_{22}|:|h_{32}|\approx
\sqrt{\frac{m_2}{m_3}}\sin\theta_{12}|1\pm
x|:\frac{1}{\sqrt{2}}|1-x^2|:\frac{1}{\sqrt{2}}|1\pm x|^2~,
\label{Yukawanormal}
\ee
where  $x=i
e^{i(\delta_1-\delta_2-\delta_3)}\sqrt{\frac{m_2}{m_3}}\cos\theta_{12}$,
and all combinations of signs are admitted. For a numerical estimate
one can take \cite{Strumia:2005tc} $\sin^2\theta_{12}\simeq 0.3$,
leading to $x\simeq 0.35 i e^{i(\delta_1-\delta_2-\delta_3)}$ and to
$|h_{12}|^2/(|h_{22}|^2+|h_{32}|^2)\sim 0.05$. In other words, the
coupling of the singlet fermion to the leptons of the first
generation is suppressed, whereas the couplings to the second and
third generations are close to each other.  

For the case of inverted hierarchy two out of four solutions are
almost degenerate and one has
\be
|h_{12}|:|h_{22}|:|h_{32}|\approx
\sqrt{\frac{1+p}{1-p}}:\frac{1}{\sqrt{2}}:\frac{1}{\sqrt{2}}~,
\label{Yukawainverted}
\ee
where $p=\pm \sin\delta_1\sin(2\theta_{12})$. Taking the same value
of $\theta_{12}$ as before, we arrive at 
$|h_{12}|^2/(|h_{22}|^2+|h_{32}|^2)\sim (0.04-25)$, depending on the
value of unknown CP-violating phase $\delta_1$. The couplings of
$N_{2,3}$ to $\mu$ and $\tau$ generations are almost identical, but
the coupling to electron and its neutrino can be enhanced or
suppressed considerably.

The corrections to relations
(\ref{Yukawanormal},\ref{Yukawainverted})  are of the order of ${\cal
O}(\epsilon)$. These predictions can be easily translated into
branching ratios of different decay modes and creation probabilities
of heavy neutrinos and can be tested if these particles are found.

%%%%%%%%%%%%%%%%%%%%%%%%%%%%%%%%%%%%%%%%%%%%%%%%%%%%%%%%%%%%%%%%%%%%%%%%
% Chapter 6
%%%%%%%%%%%%%%%%%%%%%%%%%%%%%%%%%%%%%%%%%%%%%%%%%%%%%%%%%%%%%%%%%%%%%%%%
\section{Conclusions} 
In this paper we constructed a variant of the $\nu$MSM in which the 
degeneracy of two singlet Majorana neutrinos and the lightness of the
third one is a consequence of a lepton number symmetry, broken at a
level of ${\cal O}(10^{-4})$ in both  the Yukawa and Majorana mass
sectors.  The resulting theory, as well as the more general $\nu$MSM,
explains dark matter, the baryon asymmetry of the Universe, and can
be extended to have an inflaton, being consistent with the data on
neutrino oscillations. In addition to predictions of a more general
model in the active and sterile neutrino sectors, in the almost
symmetric $\nu$MSM the couplings of two heavy neutrino mass
eigenstates to active neutrinos are almost identical, and their
flavour structure is fixed by observations of active neutrino
oscillations. Moreover, the masses of these neutral leptons are
likely to be relatively small, $M < 1$ GeV, which would allow their
search in charmed, beauty, $K$ or $\pi$-mesons or $\tau$-lepton
decays.

Besides that, the model predicts an outcome of a number of particle
physics and astrophysics experiments.  If the $\nu$MSM happens to be
a correct effective theory up to the Planck scale, the LHC or other
future accelerators will find no new particles, with the exception of
the Higgs, unless dedicated experiments to search for heavier
degenerate sterile neutrinos (like those in
\cite{Bernardi:1987ek,Astier:2001ck}), and, possibly, a light
inflaton, are performed. No new sources of CP-violation in the
hadronic sector are foreseen. The direct and indirect searches of
WIMP would give a negative result, but the dark matter particle could
be discovered in X-rays.

The bottom-up approach used in this work cannot provide any
fundamental reason for the existence of the $U(1)_L$ symmetry and
explain the required pattern for its breaking. It would be
interesting to see if the requisite symmetry and its breaking can
come from some underlying dynamics at high energy scale. This
problem, however, goes beyond the scope of the present paper.

%\newpage

%%%%%%%%%%%%%%%%%%%%%%%%%%%%%%%%%%%%%%%%%%%%%%%%%%%%%%%%%%%%%%%%%%%%%%%%
% Acknowledgments
%%%%%%%%%%%%%%%%%%%%%%%%%%%%%%%%%%%%%%%%%%%%%%%%%%%%%%%%%%%%%%%%%%%%%%%%
{\bf Acknowledgements.}
This work was supported in part by the Swiss National Science
Foundation.  I thank Dmitry Gorbunov, Mikko Laine and Igor Tkachev
for discussions and helpful remarks.
%\input{refsletter.tex}

%\input{newref.tex}

%%%%%%%%%%%%%%%%%%%%%%%%%%%%%%%%%%%%%%%%%%%%%%%%%%%%%%%%%%%%%%%%%%%%%%%%
% References
%%%%%%%%%%%%%%%%%%%%%%%%%%%%%%%%%%%%%%%%%%%%%%%%%%%%%%%%%%%%%%%%%%%%%%%%

\end{document}